\documentstyle[11pt,moriond,epsfig]{article}

\def\be{\begin{equation}}
\def\ee{\end{equation}}
\def\bea{\begin{eqnarray}}
\def\eea{\end{eqnarray}}

\begin{document}
\vspace*{4cm}
\title{Mass and Environment Drive the Evolution of Early--Type Galaxies}

\author{Sperello di Serego Alighieri}

\address{INAF -- Osservatorio Astrofisico di Arcetri,\\
Largo E. Fermi 5, 50125 Firenze, Italy}

\author{Barbara Lanzoni}

\address{INAF -- Osservatorio Astronomico di Bologna,\\
Via Ranzani 1, 40127 Bologna, Italy}

\maketitle \abstracts{
New information has recently become available on the fundamental plane
for various samples of
early-type galaxies with redshift up to 1.3, both in clusters and in the
field.
This information is reviewed and clues are derived on the evolution of
spheroids over
the last two thirds of the Universe lifetime as a function of galaxy mass
and environment, in comparison with the
predictions of
the hierarchical models of galaxy formation. Using the Universe as a
time-machine and interpreting changes in ${\cal M}/L$ ratio as age differences,
we see that the age increases with galaxy mass in all environments, cluster
galaxies with any mass are older than field galaxies with the same mass,
and the age difference between cluster and field galaxies increases with
mass. The first two results confirm those obtained with other methods, 
and are reproduced by the most
recent incarnation of the hierarchical models, while the third result is
new and appears in contrast with the predictions of these models.}

\section{Introduction}

Early--type galaxies (ETG) contain most of the visible mass in the Universe
\cite{ren06}
and are thought to reside in the highest density peaks of the underlying
dark
matter distribution. Therefore,
understanding their evolution is crucial for understanding the evolution of
galaxies and structures in general. ETG have the very interesting property
that in the 3D space of their main parameters (the effective radius $R_e$,
the central velocity dispersion
$\sigma$, and the average surface luminosity within $R_e$,  $\langle
I\rangle_e
= L/2\pi R_e^2$) they concentrate on a plane, therefore called the
fundamental plane (FP \cite{djo87,dre87}). This implies a striking
regularity in their structure and stellar population \cite{ren93}, which
calls for a uniform process of formation and evolution. Studies of the FP
at high redshift allow for a check on the persistence of this regularity
back in time and offer the possibility to derive more stringent constraints
on the formation history of ETG, using the Universe as a time-machine.

We have therefore made a uniform comparison of the best data available on
the FP at $z$$\sim$1, the highest redshift for which these data are
currently available, and, by interpreting the differences in ${\cal M}/L$ ratio
directly derived from the FP parameters as age differences, we study how
the ages of ETG depend on both galaxy mass and environment.
We assume a flat Universe with $\Omega_m=0.3$,
$\Omega_{\Lambda}=0.7$, and $H_0=70$ km s$^{-1} {\rm Mpc}^{-1}$, and
we use magnitudes based on the Vega system.

\section{The data on the FP at $z$$\sim$1}

In selecting the data on the FP at $z$$\sim$1 we pay particular attention
to the completeness of the samples to faint optical magnitudes, because the
galaxies with correspondingly small masses are those likely to show
``downsizing'' effects, i.e. the later and/or longer lasting formation of 
smaller mass galaxies \cite{cow96}, and differences with environment. 
In fact it has been shown that massive galaxies in the field behave similarly
to massive galaxies in the clusters \cite{dsa05}. We also consistently use
data in the rest--frame B band. In the field the only
ETG galaxy sample complete to faint magnitudes at $z$$\sim$1, for which FP 
studies have been made, is the one derived from the K20 survey \cite{dsa05},
which is complete down to $M_B=-20.0$, and covers the redshift range
$0.88<z<1.3$ with 15 ETG over a total sky solid angle of 52 arcmin$^2$. We
have used also the sample of Treu et al.\cite{tre05}, which, although not complete, 
contains 24 ETG in the same redshift range over 160 arcmin$^2$ in the
GOODS-N area, some of which are as faint as the faintest galaxies of di
Serego Alighieri et al.\cite{dsa05}. We have also examined the data
presented by van der Wel et al.\cite{van05}, who have studied the FP on a
sample of 27 field ETG in the range $0.6<z<1.15$ in two Southern fields,
but eventually were forced not to include them in our analysis
by the impossibility of understanding how they have selected the 13 primary
ETG plotted in their figures from the 27 galaxies in their sample.

Until very recently there was no
sample of cluster ETG at high redshift with FP information and complete to faint
magnitudes \cite{van03}. This is probably due to an observational selection effect,
since on the clusters the slits of a single mask in a MOS spectrograph are 
easily filled with bright galaxies, and none is left for the fainter ones.
We were very lucky that shortly before this Conference a FP study
\cite{jor06} of two
clusters at $z=0.835$ and $z=0.892$, complete down to $M_B=-19.8$, has been
published, and, although the data for the individual galaxies
are not yet published, Inger J\o rgensen has kindly made them available to us
for the present analysis.

We have uniformly analized these three sets of data \cite{dsa05,tre05,jor06}
and we show in
Fig. 1 the corresponding edge--on FP, compared to a recent analysis of the
Coma cluster ETG \cite{jor06}, where galaxies with 
${\cal M} < 10^{10.3}{\cal M}_{\odot}$ and emission--line galaxies have
been excluded.

\begin{figure}
\begin{center}
\psfig{figure=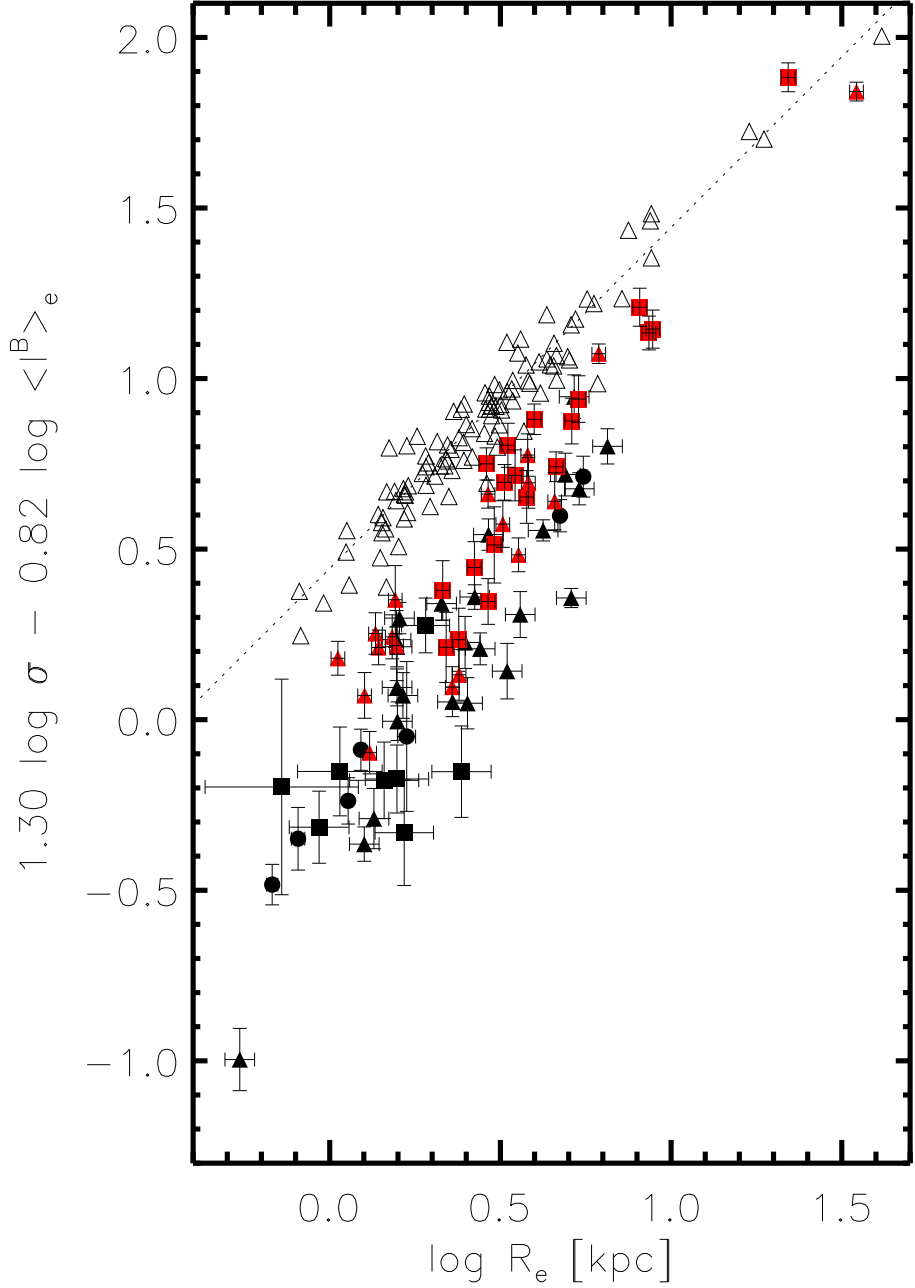,height=18.5cm}
\caption{The edge--on Fundamental Plane for local ETG in the Coma
Cluster (empty triangles), for field ETG at $z$$\sim$1 from
the K20 survey both for the CDFS field (filled black circles)
and for the Q0055 field (filled black squares), for field ETG at $z$$\sim$1
in the GOODS area (filled black triangles), and for the ETG in
two
clusters at $z=0.835$ (red filled squares) and at $z=0.892$ (red
filled
triangles). The dashed line is the best fit plane to the Coma cluster
galaxies.
Compared to the local one, the FP at high redshift is offset and rotated in
all environments.
\label{fig:FP}}
\end{center}
\end{figure}

It is evident from Fig. 1 that the evolution of the FP consists both of a
vertical shift and of a rotation, in the sense that the FP at $z$$\sim$1
appears steeper that the local one, both in the field and in the clusters.
As has been shown before \cite{dsa05}, the change in the FP tilt is a necessary manifestation
of ``downsizing''.
Notwithstanding this evolution, the FP has a remarkably small scatter also
at $z$$\sim$1.

\section{Deriving ages from the FP parameters}

It is customary and straightforward to interpret the evolution of the FP as
changes in the ${\cal M}/L$ ratio. In fact the FP parameters can be used to estimate
the dynamical mass of the galaxies. For instance, from the Virial theorem
and assuming $R^{1/4}$ homology, the mass is given by \cite{mic80}:

\begin{equation}
{\cal M} = 5R_e\sigma^2/G.
\end{equation}

Then the ${\cal M}/L_B$ ratio can be obtained for each high redshift galaxy and
compared to that of local galaxies with the same mass, as shown in Fig. 2.

\begin{figure}
\begin{center}
\psfig{figure=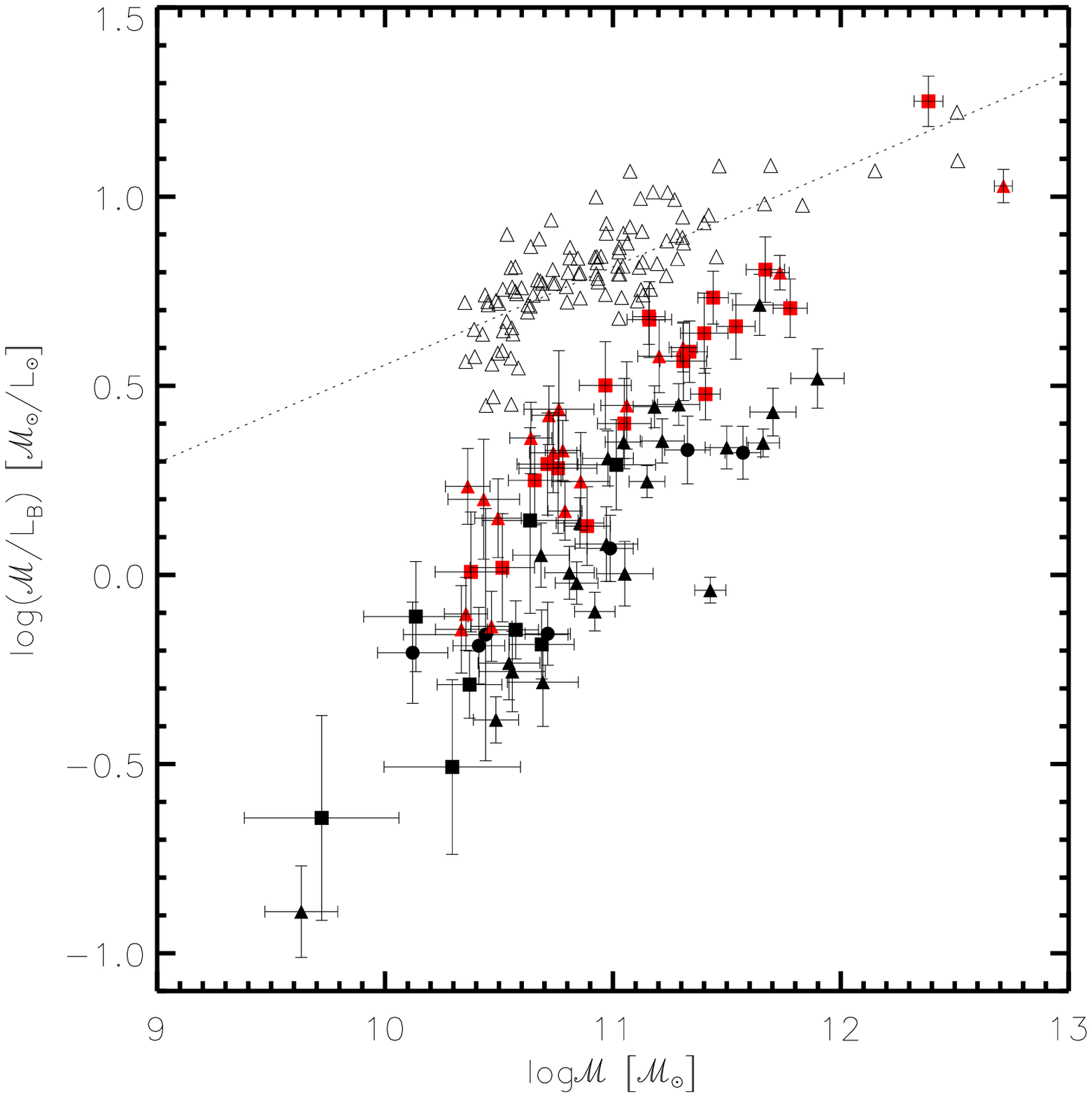,height=10cm}
\caption{The ${\cal M}/L$ ratio in the B-band as a function of the galaxy mass
for the ETG samples given in Figure 1 (same symbols). The dotted line is
a fit to the Coma ETG. The changes in ${\cal M}/L_B$ from
high redshift to today decrease with galaxy mass in all environments and are
larger in the field than in the clusters.
\label{fig:MtoL}}
\end{center}
\end{figure}

Evidently the difference in the ${\cal M}/L$ ratio between a local ETG and one at 
$z$$\sim$1 with the same mass is larger for smaller mass galaxies, both in
the field and in the clusters, and seems smaller in the cluster than in the
field. However the redshift of the two clusters is
smaller that the average redshift of the field galaxies: therefore a deeper
analysis is necessary to ensure that the differences with environment are
not due just to redshift differences. Usually this is achieved by taking
for each high redshift ETG the logarythmic difference in ${\cal M}/L$ ratio ($\Delta
log({\cal M}/L_B$)) from
that of a local galaxy with the same mass and comparing it to the straight
line fit  to the $\Delta
log({\cal M}/L_B)$ of ETG with ${\cal M}>10^{11} {\cal M}_{\odot}$ in clusters with redshift
ranging from 0 to 1.3 \cite{van03}. However this analysis is unsatisfactory,
because it is not clear why the massive cluster galaxies should be taken as a
reference, and because it excludes the lower mass cluster galaxies from the
analysis. We have therefore devised a different analysis with the aim of
studying how the formation history of ETG depends on galaxy mass and
environment. We interpret the changes in ${\cal M}/L$ ration as differences in age.
Maraston \cite{mar05} has developed simple stellar population synthesis models 
that allow one to estimate the ${\cal M}/L$ ratio as a function of age (Fig. 3).

\begin{figure}
\begin{center}
\psfig{figure=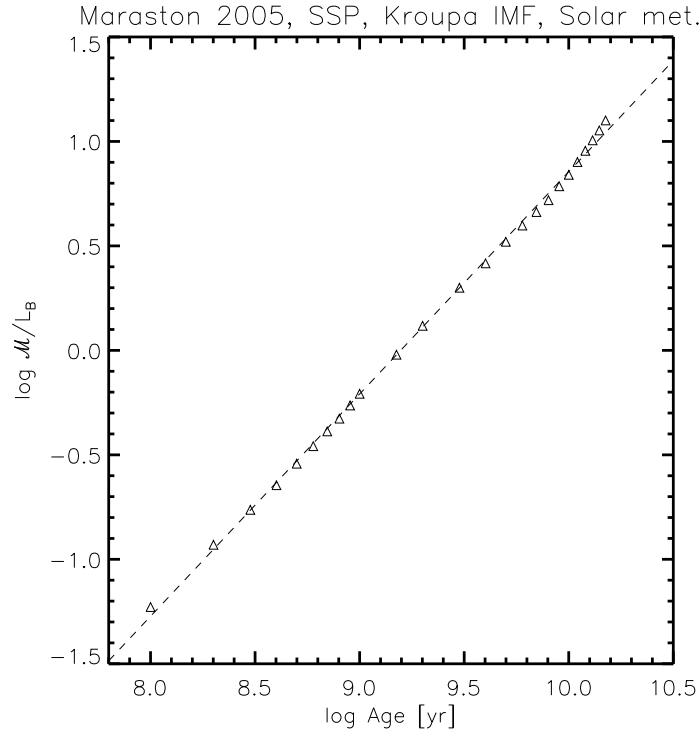,height=10cm}
\caption{The relationship between age and ${\cal M}/L$ ratio obtained by Maraston (2005)
for a simple stellar population with solar metallicity. The dashed line shows
the linear fit that we have used.
\label{fig:MarMod}}
\end{center}
\end{figure}

By fitting this relation, e.g. for solar metallicity, nd for a Kroupa
\cite{kro01} we can therefore get 
the age (at $z=0$) of each galaxy from the observed $\Delta
log({\cal M}/L_B)$ for our assumed cosmology. The results are shown in Figure 4,
compared to the median ages derived from a semianalytic model of
hierarchical galaxy evolution \cite{del06}.

\begin{figure}
\begin{center}
\psfig{figure=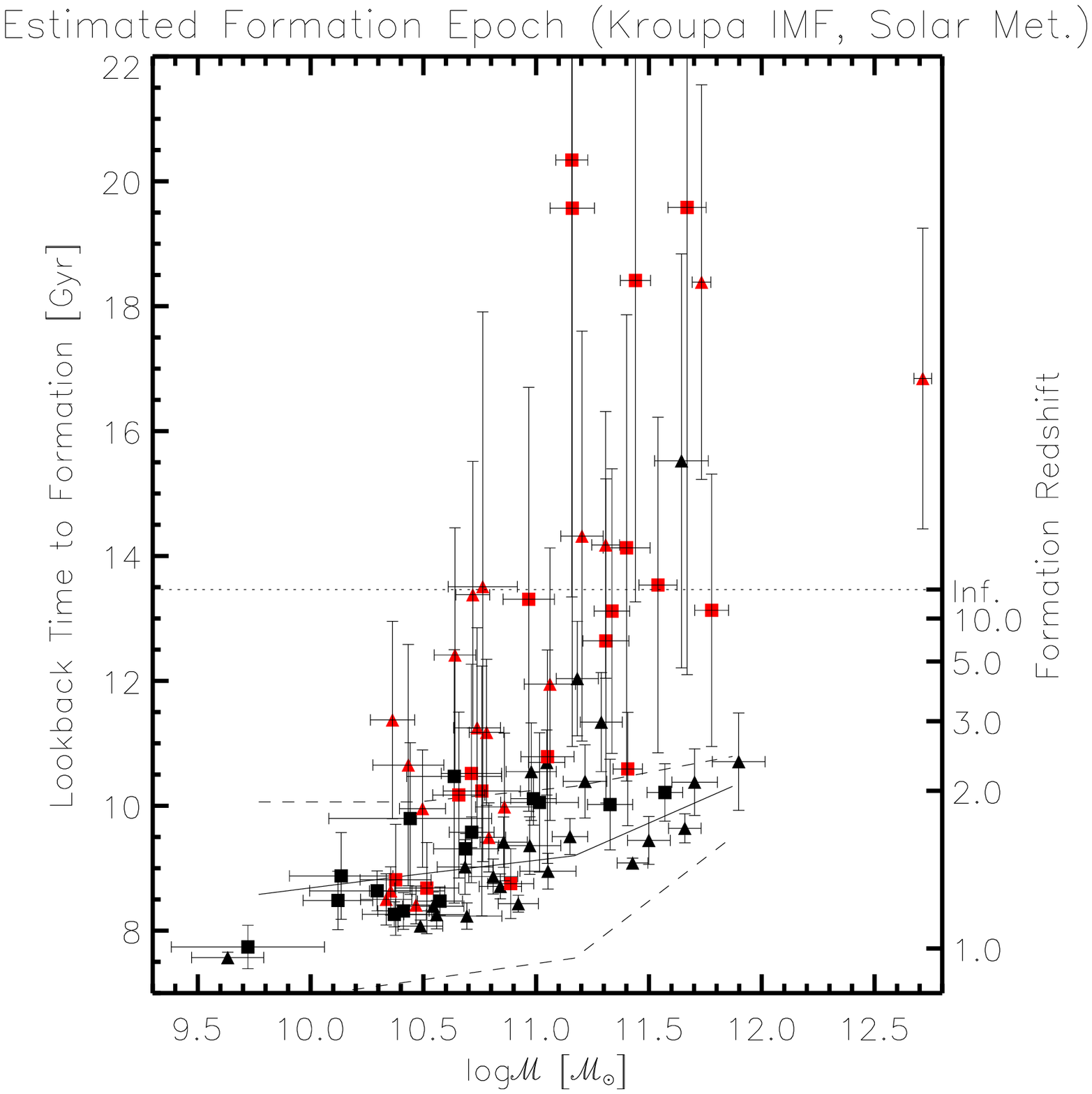,height=16cm}
\caption{The formation epoch of the ETG from the samples given in Figure 1
(same symbols), evaluated as explained in the text. The continuous line
shows the median model ages obtained by De Lucia et al. (2006) from a semianalytic
model of hierarchical galaxy formation, while the dashed lines are their
upper and lower quartiles.
More massive ETG form earlier in all environments, and
cluster ETG are older than those in the field at all masses, while their
age
difference appears to increase with galaxy mass. We stress that, although
the absolute ages that we derive are somewhat model dependent and reach
a few absurdly high values, the trends of age differences between high redshift
and local ETG, and between galaxies with different masses and in different
environments are much more robust.
\label{fig:Ages}}
\end{center}
\end{figure}

ETG age increases with mass (downsizing) both in the field and in the
clusters. Field galaxies are younger than cluster galaxies with the same
mass. The age difference between cluster and field galaxies with the same
mass increases with mass. The first two results have been obtained before
by other means and are reproduced by the most recent incarnation of the
hierachical models. The third result is new and appears to go in the
opposite way as predicted by the hierarchical models (see fig. 1 of De Lucia
et al. \cite{del06}).

Note that after the Conference we have refined our estimate of galaxy ages,
by evaluating the metallicity for each ETG from the velocity dispersion and
then direclty using the population synthesis model appropriate for the derived
metallicity \cite{mar05} without any fitting involved. We have also
analysed selection effects. The resulting ages are 
very similar to those presented here and our conclusions are not affected 
\cite{dis06}.

\section*{Acknowledgments}

We would like to thank Inger J\o rgensen for allowing us to use her still
unpublished data, and Claudia Maraston and Alvio Renzini for useful
suggestions and discussions.

\end{document}